\documentclass[superscriptaddress,aps,preprintnumbers,amsmath,showpacs,amssymb,prd,nofootinbib,reprint]{revtex4-1}

\usepackage[dvipdfmx]{graphicx}
\usepackage{amsfonts}
\usepackage[mathscr]{eucal}
\usepackage{ascmac}
\usepackage{dcolumn}
\usepackage{bm}
\usepackage[colorlinks=true,linkcolor=blue,citecolor=blue]{hyperref}
\usepackage{hyperref}
\usepackage{color}

\begin{document}


\newcommand{\vev}[1]{ \left\langle {#1} \right\rangle }
\newcommand{\bra}[1]{ \langle {#1} | }
\newcommand{\ket}[1]{ | {#1} \rangle }
\newcommand{\EV}{ \ {\rm eV} }
\newcommand{\KEV}{ \ {\rm keV} }
\newcommand{\MEV}{\  {\rm MeV} }
\newcommand{\GEV}{\  {\rm GeV} }
\newcommand{\TEV}{\  {\rm TeV} }
\newcommand{\1}{\mbox{1}\hspace{-0.25em}\mbox{l}}
\newcommand{\Red}[1]{{\color{red} {#1}}}

\newcommand{\lmk}{\left(}  
\newcommand{\rmk}{\right)}
\newcommand{\lkk}{\left[}  
\newcommand{\rkk}{\right]}
\newcommand{\lhk}{\left \{ }  
\newcommand{\rhk}{\right \} }
\newcommand{\del}{\partial}  
\newcommand{\la}{\left\langle} 
\newcommand{\ra}{\right\rangle}
\newcommand{\half}{\frac{1}{2}}

\newcommand{\bea}{\begin{array}}
\newcommand{\eea}{\end{array}}
\newcommand{\beq}{\begin{eqnarray}}
\newcommand{\eeq}{\end{eqnarray}}

\newcommand{\dd}{\mathrm{d}}
\newcommand{\Mpl}{M_{\rm Pl}}
\newcommand{\mg}{m_{3/2}}
\newcommand{\abs}[1]{\left\vert {#1} \right\vert}
\newcommand{\mphi}{m_{\phi}}
\newcommand{\Hz}{\ {\rm Hz}}
\newcommand{\for}{\quad \text{for }}
\newcommand{\Min}{\text{Min}}
\newcommand{\Max}{\text{Max}}
\newcommand{\Kahler}{K\"{a}hler }
\newcommand{\cphi}{\varphi}
\newcommand{\Tr}{\text{Tr}}
\newcommand{\diag}{{\rm diag}}

\newcommand{\SUf}{SU(3)_{\rm f}}
\newcommand{\Upq}{U(1)_{\rm PQ}}
\newcommand{\Zpq}{Z^{\rm PQ}_3}
\newcommand{\Cpq}{C_{\rm PQ}}
\newcommand{\ubar}{u^c}
\newcommand{\dbar}{d^c}
\newcommand{\ebar}{e^c}
\newcommand{\nubar}{\nu^c}
\newcommand{\Ndw}{N_{\rm DW}}
\newcommand{\Fpq}{F_{\rm PQ}}
\newcommand{\fpq}{v_{\rm PQ}}
\newcommand{\Br}{{\rm Br}}
\newcommand{\Lag}{\mathcal{L}}
\newcommand{\Lqcd}{\Lambda_{\rm QCD}}
\newcommand{\cm}{{\rm \ cm}}


\preprint{
IPMU 15-0185; 
DESY 15-189
}

\title{
Cosmologically safe QCD axion without fine-tuning 
}

\author{
Masaki Yamada
}
\affiliation{Kavli IPMU (WPI), UTIAS, 
The University of Tokyo, 
Kashiwa, Chiba 277-8583, Japan}
\affiliation{Institute for Cosmic Ray Research, 
The University of Tokyo, 
Kashiwa, Chiba 277-8582, Japan}
\affiliation{
Deutsches Elektronen-Synchrotron DESY, 
22607 Hamburg, Germany
}

\author{
Tsutomu T. Yanagida
}
\affiliation{Kavli IPMU (WPI), UTIAS, 
The University of Tokyo, 
Kashiwa, Chiba 277-8583, Japan}

\author{
Kazuya Yonekura
}
\affiliation{Kavli IPMU (WPI), UTIAS, 
The University of Tokyo, 
Kashiwa, Chiba 277-8583, Japan}

\date{\today}

\begin{abstract} 
Although QCD axion models are widely studied as solutions to the strong CP problem, 
they generically confront severe fine-tuning problems to guarantee the anomalous PQ symmetry. 
In this letter, 
we propose a simple QCD axion model without any fine-tunings. 
We introduce an extra dimension 
and a pair of extra quarks living on two branes separately, 
which is also charged under a bulk Abelian gauge symmetry. 
We assume a monopole condensation on our brane at an intermediate scale, 
which implies that 
the extra quarks develop the chiral symmetry breaking and the PQ symmetry is broken. 
In contrast to the original Kim's model, 
our model explains the origin of the PQ symmetry thanks to the extra dimension 
and avoids the cosmological domain wall problem because of the chiral symmetry breaking in the Abelian gauge theory.

\end{abstract}

\maketitle


\section{Introduction
\label{sec:introduction}}

The null result of neutron electric dipole moment 
puts an upper bound on the magnitude of the strong CP phase: 
$\theta_{\rm QCD} \lesssim 10^{-(10-11)}$~\cite{Baker:2006ts}. 
Such a small fundamental parameter is an outstanding mystery in particle physics, 
known as the strong CP problem. 
Peccei and Quinn have proposed a mechanism to explain its smallness 
by introducing an anomalous global Abelian symmetry, called the PQ symmetry~\cite{Peccei:1977hh, Peccei:1977ur}. 
The PQ symmetry is assumed to be broken spontaneously at an intermediate scale, 
which 
predicts a pseudo-Nambu-Goldstone (NG) boson, called an axion~\cite{Weinberg:1977ma}. 
The axion obtains a periodic potential due to the nonperturbative effect 
below the QCD scale 
and its vacuum expectation value (VEV) cancels the unwanted CP phase~\cite{'tHooft:1976up, 'tHooft:1976fv}. 
Here, 
in order to suppress the effective CP phase sufficiently, 
the PQ symmetry breaking effects should be suppressed precisely. 
In general, however, 
quantum gravity effects may induce PQ symmetry breaking terms, 
so that severe fine-tunings may be required to explain the smallness of the strong CP phase~\cite{
Giddings:1988cx, Coleman:1988tj, Gilbert:1989nq, Banks:2010zn, Kamionkowski:1992mf, 
Ghigna:1992iv, Dobrescu:1996jp}. 

In Refs.~\cite{Izawa:2002qk, Izawa:2004bi}, 
it has been revealed that a model with a flat extra dimension 
can explain the origin of the PQ symmetry without any fine-tunings.%
\footnote{
See Refs.~\cite{Holman:1992us, Cheng:2001ys, Barr:2001vh, Harigaya:2015soa} for other mechanisms 
to explain the origin of PQ symmetry. 
}
The 5D manifold is $R^4 \times S^1 / Z_2$ 
and two branes are located at the fixed points in the $S^1 / Z_2$ orbifold. 
Each extra quark-antiquark pair lives on branes separately, 
which results in suppression of PQ symmetry breaking operators. 
Introducing a bulk strong gauge interaction under which the extra quarks are charged, 
one can consider 
the dynamical PQ symmetry breaking scenario~\cite{Kim:1984pt}. 
A composite NG boson is identified with axion 
and the strong CP problem is solved without fine-tunings. 
Unfortunately, however, 
there are cosmological difficulties in this scenario. 
If the PQ symmetry is dynamically broken before primordial inflation 
and is never restored after inflation, 
the axion acquires quantum fluctuations during inflation 
and predicts axion isocurvature perturbation, 
which is severely constrained by the observations of CMB temperature fluctuations~\cite{Axenides:1983hj, 
Seckel:1985tj, Turner:1990uz}.%
\footnote{
In Ref.~\cite{Kawasaki:2015lea}
it has been proposed a scenario to solve the 
axion isocurvature problem 
in the case that the PQ symmetry is broken before inflation. 
}
On the other hand, 
if the PQ symmetry is restored 
and then dynamically broken after inflation, 
e.g., in a high reheating temperature scenario, 
domain walls form 
at the QCD phase transition~\cite{Zeldovich:1974uw, Sikivie:1982qv}. 
If the bulk strong interaction is $SU(N)_H$ gauge interaction, 
the domain wall number is $N$. 
In this case, 
domain walls are stable 
and soon dominate the energy density of the Universe 
after the QCD phase transition. 
Since the resulting Universe is highly inhomogeneous, 
this scenario is excluded. 
This is known as the axion domain wall problem.

Here, let us assume that 
the chiral symmetry breaking is induced by a bulk $U(1)_H$ gauge interaction 
rather than $SU(N)_H$. 
In this case, 
the PQ symmetry is dynamically broken completely at the QCD phase transition, 
so that the domain wall number is unity. 
Since these domain walls are unstable due to their tension, 
there is no domain wall problem in this case. 
Therefore, 
we can safely consider the scenario that the PQ symmetry is broken after inflation.

The next question is 
how the chiral symmetry breaking occurs in the Abelian gauge theory. 
Actually, 
there are many works related to the confinement and chiral symmetry breaking 
in Abelian-like gauge theories 
to reveal the QCD confinement and chiral symmetry breaking. 
't~Hooft conjectured that 
long-distance physics could be realized only by an Abelian degrees of freedom in QCD, 
which is called as Abelian dominance~\cite{'tHooft:1981ht}. 
The lattice QCD studies in fact reveal that the confinement and chiral symmetry breaking 
may occur by an Abelian-projected field in the maximally Abelian gauge~\cite{Stack:1994wm, Miyamura:1995xn, Woloshyn:1994rv}, 
where a monopole current plays an important role~\cite{Kronfeld:1987vd, Kronfeld:1987ri}. 
These studies may imply that 
electron confining and chiral symmetry breaking can occur in an Abelian gauge theory. 
The t'~Hooft's conjecture stands on 
an Abelian theory with a magnetic superconductor for confinement, 
where a physical string arises between an electron and anti-electron 
because of a monopole condensation~\cite{Nambu:1974zg}. 
Motivated by these studies, 
in this letter we introduce a monopole on our brane 
and assume the chiral symmetry breaking of electrons (extra quarks) by the condensation of monopole. 

Our model with a monopole and electrons may also be supported by 
Refs.~\cite{Seiberg:1994rs, Seiberg:1994aj, Argyres:1995jj, Argyres:1995xn}, 
where they have discovered 
UV complete theories with monopoles and electrons. 
They consider a $SU(2)$ gauge theory with at most three flavour electrons 
in $N=2$ supersymmetry. 
In a $N=2$ supersymmetric multiplet, 
$SU(2)$ gluons are accompanied by Dirac fermions and complex scalars in the adjoint representation of the gauge group. 
The moduli space of this theory has a branch called the Coulomb branch, 
on which 
the complex scalar component of gauge multiplet has a nonzero VEV 
and 
the $SU(2)$ gauge symmetry is spontaneously broken down to $U(1)$. 
As a result, a monopole appears in the low energy effective theory. 
There is a singular point on the Coulomb branch 
at which the monopole becomes massless. 
In addition, once we introduce an equal mass for electrons, 
there is another singularity on the moduli space 
at which the electrons are massless. 
When these singular points coincide with each other, 
the low energy effective theory contains a massless monopole and massless electrons 
with an Abelian gauge field~\cite{Argyres:1995jj, Argyres:1995xn}. 
This theory motivates us to consider a model with a monopole and electrons in an Abelian gauge theory. 
Although we do not have a UV complete theory like their $N=2$ supersymmetric theory, 
we consider the model in a bottom-up approach to cosmological problems.%
\footnote{
Although one might wonder that the electric coupling constant blows up in the UV limit in our model, 
we expect that it asymptotically approaches a UV fixed point as discussed in Ref.~\cite{Argyres:1995jj, Argyres:1995xn}. 
}

\section{chiral symmetry breaking in $U(1)$ gauge theory 
\label{sec:chiral symmetry breaking}}

We consider a non-supersymmetric $U(1)_H$ gauge theory in this letter. 
(However, its supersymmetric extension will be strait-forward.) 
We introduce a scalar monopole $\phi$ and fermionic electrons, 
which are charged under a $U(1)_H$ gauge symmetry. 
We denote the unit electric and monopole charges as $e$ and $g$, respectively, 
which are related by the Dirac quantization condition: 
\beq
 e g = \frac{4\pi}{2} n, 
\eeq
where $n= 0, \pm 1, \pm 2, \dots$. 
We assume that 
the $U(1)_H$ gauge symmetry is spontaneously broken 
by the condensation of the monopole: 
\beq
 \la \phi \ra \equiv v, 
\eeq
and the $U(1)_H$ gauge boson acquires an effective mass of $m_v = g v$. 
The spontaneous symmetry breaking (SSB) of $U(1)_H$ symmetry implies 
the formation of 
cosmic strings. 
In fact, 
each electron and positron pair is connected with a cosmic string~\cite{Nambu:1974zg}. 
The tension of the string is calculated as 
\beq
 \mu = \frac{e^2}{8 \pi} m_v^2 \log \lmk \frac{m_m^2}{m_v^2} + 1 \rmk, 
\eeq
where $m_m$ is the mass of the monopole. 
Note that the tension is almost independent of $e$ and $g$ 
because of the Dirac quantization condition. 
Since each electron and positron pair is connected by the string, 
they are confined.

Here, let us consider a low energy effective theory 
below the confinement scale. 
Since we consider the Abelian gauge theory with fermionic electrons and positrons whose charges are unity, 
there are only boson states in the low energy. 
This implies that 
the theory cannot satisfy 
the t'Hooft anomaly matching about the chiral symmetry between the low and high energy scales 
unless the chiral symmetry is dynamically broken. 
Thus, 
we assume that the chiral symmetry is dynamically broken below the confinement scale.

\section{origin of PQ symmetry
\label{sec:PQ symmetry}}

Now we consider a theory with an extra dimension 
where $5D$ manifold is $R^4 \times S/Z_2$. 
Two branes are located at the fixed points in the $S^1 / Z_2$ orbifold, 
one of which the monopole as well as the SM particles live on. 
We introduce $N_F$ ($\ge 4$) pairs of electrons $Q^i$ and $\bar{Q}^i$ 
and put them separately on our brane and the other brane~\cite{Izawa:2002qk, Izawa:2004bi}. 
These electrons are charged under the hidden Abelian gauge symmetry $U(1)_H$ 
as shown in Table~\ref{table1}. 
The $U(1)_H$ gauge field as well as the SM gauge fields propagate in the bulk. 
The first three $Q^i$ ($\bar{Q}^i$) of ``flavour'' indices 
$i=1,2,3$ transform under the fundamental (anti-fundamental) representation of $SU(3)_c$, 
so that we call them as an extra quark (anti-quark). 
Since 
direct contact interactions between $Q$ and $\bar{Q}$ are suppressed exponentially 
due to the separation in the extra dimension~\cite{Izawa:2002qk, Izawa:2004bi}, 
there is an approximate chiral symmetry in our model. 

As explained in the previous section, 
we consider that 
the chiral symmetry is dynamically broken 
at an intermediate scale $f_a$ ($\approx v$): 
\beq
 \la Q^i \bar{Q}^j \ra \simeq f_a^3 \delta^{ij}. 
\eeq
If the QCD interaction was turned off, 
there is the $SU(N_F)_L \times SU(N_F)_R \times U(1)_A$ flavour symmetry 
before the chiral symmetry breaking.%
\footnote{
Note that the vector symmetry is just the $U(1)_H$ gauge symmetry. 
}
It is broken down to $SU(N_F)_V$ by the chiral symmetry breaking 
and $(N_F^2 - 1)$ NG bosons may arise in the low energy effective theory. 
However, 
the flavour symmetry is explicitly broken by $SU(3)_c$ gauge interactions, 
so that $SU(3)_c$ charged NG bosons obtain masses via $SU(3)_c$ radiative corrections. 
The $U(1)_A [ U(1)_H]^2$ anomaly induces an effective mass to the pseudo-NG boson 
associated with the $U(1)_A$ symmetry.%
\footnote{
The pseudo-NG boson corresponding to the axial anomaly 
may obtain an effective mass 
by the Witten effect 
in the presence of monopole~\cite{Witten:1979ey, Fischler:1983sc}. 
}
Finally, 
the QCD axion can be identified with a pseudo-NG boson 
associated with 
the linear combination of axial symmetries shown in Table~\ref{table1} as $U(1)_{\rm PQ}$~\cite{Kim:1984pt}. 
To sum up, 
there are $(N_F - 3)^2 - 1$ massless NG bosons 
and the axion 
in the low energy effective theory. 
Note that when $N_F = 4$, 
which is the minimal number of allowed values of $N_F$ in our model, 
there is only the axion in the low energy effective theory.

\begin{table}\begin{center}
{\renewcommand\arraystretch{1.5}
\begin{tabular}{|c|p{1.6cm}|p{1.6cm}|p{1.6cm}|p{1.6cm}|}
  \hline
    & \hfil $Q^{i(=1,2,3)}$ \hfil & \hfil $\bar{Q}^{i(=1,2,3)}$ \hfil & \hfil $Q^{i (\ge 4)}$ \hfil & \hfil $\bar{Q}^{i (\ge 4)}$ \hfil \\
  \hline
  \hfil $SU(3)_c$ \hfil & \hfil {\bf 3} \hfil & \hfil ${\bf 3}^*$ \hfil & \hfil {\bf 1} \hfil & \hfil {\bf 1} \hfil \\
  \hline
  \hfil $U(1)_{\rm H}$ \hfil & \hfil 1 \hfil & \hfil $-1$ \hfil & \hfil 1 \hfil & \hfil $-1$ \hfil \\
  \hline
  \hfil $U(1)_{\rm PQ}$ \hfil & \hfil 0 \hfil & \hfil 1 \hfil & \hfil 0 \hfil & \hfil $-\frac{3}{N_F -3}$ \hfil \\
\hline
\end{tabular}
}
\end{center}
\caption{Charge assignment for matter fields.
\label{table1}}
\end{table}

There are two advantages in this model. 
First, 
the PQ symmetry is completely broken by non-perturbative effect 
after the QCD phase transition, 
so that the domain wall number is unity. 
Since domain walls are unstable due to their tension in this case, 
there is no domain wall problem. 
Secondly, 
the $U(1)_H$ symmetry forbids operators containing either $Q$ or $\bar{Q}$, 
which arise and may be problematic in the original model with a $SU(N)_H$ gauge symmetry~\cite{Izawa:2002qk, Izawa:2004bi}. 
This also guarantees the precision of the PQ symmetry. 

Here we comment on hadron states of electrons. 
There is no baryon state in the low energy because the confinement of electrons is induced by the Abelian 
$U(1)_H$ interaction. 
There are heavy meson states in the low energy. 
In particular, composite states like $\pi^i \equiv (Q^{i} \bar{Q}^4)$ (hereafter, we take $i=1,2,3$) 
may be stable heavy mesons 
and their energy density may overclose the Universe. 
We can make them unstable in the following way~\cite{Izawa:2004bi}.%
\footnote{
Instead, one can assume that these heavy mesons are washed out by the primordial inflation 
and are never produced by reheating process. 
This is the case if the reheating temperature is much lower than the dynamical scale. 
}
First, 
we assume that the fields $Q^i$ and $\bar{Q}^i$ have 
the same SM gauge charge with the right-handed down quark and its complex conjugate, respectively, 
which is also motivated by grand unified theories. 
We also introduce a heavy scalar $\chi^i$ which 
lives on our brane (like $Q^i$) 
and has the same SM gauge charge with $\bar{Q}^i$.%
\footnote{
In the supersymmetric extension we can identify the field $\chi^i$ with a scalar down-type quark
as pointed out in Ref.~\cite{Izawa:2004bi}. The scalar quark can decay quickly
if the SM gauginos are lighter than the scalar quark. 
}
Now we can write 
dimension five interactions of $((Q^4)^\dagger \gamma^\mu Q^i) D_\mu \chi_i + c.c.$, 
where $D_\mu$ is the covariant derivative. 
After the chiral symmetry breaking 
these interactions become the operators of $D_\mu \pi^i D_\mu \chi_i +c.c.$, 
which imply the kinetic mixing between $\pi^i$ and $\chi^i$. 
Once we make the field $\chi^i$ decay into SM particles 
via Yukawa interactions of $\epsilon_{i j k} \chi^i q_L^j q_L^k + c.c.$, 
where $q_L$ is the left-handed quark, 
the heavy meson $\pi^i$ also decays via the kinetic mixing.%
\footnote{
One might wonder if the Yukawa interactions lead to dangerous proton decay. 
However, 
if the mass of $\chi^i$ is of order $10^{10} \GEV$ and the Yukawa coupling constants 
to the first family quarks and leptons are $\mathcal{O}(10^{-6})$, 
the proton lifetime is consistent with the present lower bound. 
If $\chi^i$ couples to the third family quarks and leptons with Yukawa couplings of $\mathcal{O}(1)$, 
the decay temperature of $\pi^i$ is as large as $10^5 \GEV$ 
and its energy density never dominate the Universe. 
}
The above interactions never violate the PQ symmetry defined in Table~\ref{table1} 
because we can introduce the dimension five interactions only on our brane, where the SM particles, 
$Q^i$, and $\chi^i$ live.

\section{predictions 
\label{sec:predictions}}

Since the PQ symmetry is dynamically broken after inflation, 
a cosmic string and domain wall system arises after the QCD phase transition. 
In our model, the domain wall number is unity, 
so that the system is short-lived and disappears soon after it forms. 
Axions are generated from the decay of these topological defects as well as 
from the usual misalignment mechanism~\cite{Preskill:1982cy, Abbott:1982af, Dine:1982ah}. 
Here, we should take the average in the initial angle of axion 
because it randomly distributes in the phase space. 
To sum up, 
the axion abundance is given by~\cite{Kawasaki:2014sqa}
\beq
 \Omega_a h^2 \simeq (1.7 \pm 0.4 ) \times 10^{-2} 
 \times \lmk \frac{f_a}{10^{10} \GEV} \rmk^{1.19}. 
\eeq
The observed DM abundance implies that 
the axion decay constant is given by 
\beq
 f_a \simeq (4.2 - 6.5) \times 10^{10} \GEV. 
\eeq
Note that this decay constant corresponds to the axion mass of 
\beq
 m_a \simeq (0.9 - 1.4) \times 10^{-4} \EV. 
\eeq
Notably, 
the Axion Dark Matter Experiment (ADMX) will be improved 
by the use of higher harmonic ports (ADMX-HF) 
and will cover the axion mass range of $16-160 \mu \EV$~\cite{vanBibber:2013ssa}.

There are $(N_F - 3)^2 - 1$ NG bosons as well as the axion 
in the low energy effective theory. 
Since the PQ symmetry is broken after inflation, 
they are in the thermal equilibrium 
and then decouple from the thermal plasma 
after the PQ symmetry breaking~\cite{Turner:1986tb, Salvio:2013iaa}. 
Their abundance is determined by the conservation of entropy density 
and is conventionally expressed by the effective neutrino number as~\cite{Nakayama:2010vs, Weinberg:2013kea, Kawasaki:2015ofa}
\beq
 N_{\rm eff} \simeq N_{\rm eff}^{(SM)} + 0.027 \times (N_F - 3)^2, 
\eeq
where $N_{\rm eff}^{(SM)}$ ($\simeq 3.046$) is the SM prediction. 
The present constraint is $N_{\rm eff} = 2.99 \pm 0.39$ ($95\%$ C.L.)~\cite{Planck:2015xua}, 
so that $N_F$ has to be smaller than or equal to $6$. 
The ground-based Stage-IV CMB polarization experiment CMB-S4 will measure 
the effective neutrino number with a precision of $\Delta N_{\rm eff} = 0.0156$ 
within one sigma level~\cite{Abazajian:2013oma, Wu:2014hta}. 
If the number of flavours would be measured via the observation of the effective neutrino number, 
it would be a remarkable evidence of our model. 
Unfortunately, however, 
our model 
cannot distinguish with the KSVZ axion model 
in the case of $N_f =4$~\cite{Kim:1979if}.

\section{discussion and conclusions
\label{sec:conclusion}}

We have proposed a QCD axion model 
where 
the PQ symmetry accidentally arises due to the separation of extra quark and anti-quark pairs in an extra dimension. 
In contrast to the original theory proposed in Ref.~\cite{Izawa:2002qk, Izawa:2004bi}, 
we avoid the axion domain wall problem 
by assuming a chiral symmetry breaking in an Abelian gauge theory. 
In fact, there are some proposals that 
Abelian gauge theories develop confinement and chiral symmetry breaking, 
including the 't~Hooft's conjecture of Abelian dominance~\cite{'tHooft:1981ht}, 
which is supported by lattice QCD~\cite{Stack:1994wm, Miyamura:1995xn, Woloshyn:1994rv}, 
and the confinement theory by monopole condensation in an Abelian theory~\cite{Nambu:1974zg}. 
There are also UV complete models 
with electrons and monopoles in an Abelian gauge theory in $N=2$ supersymmetry~\cite{Seiberg:1994rs, Seiberg:1994aj, Argyres:1995jj, Argyres:1995xn}. 
Motivated by these studies, 
we consider an Abelian gauge theory 
with chiral symmetry breaking of electrons 
triggered by a monopole condensation. 
Since we consider the Abelian theory rather than non-Abelian theories, 
the PQ symmetry is completely broken at the QCD phase transition 
and domain walls disappear soon after they form at that time.

Finally, we comment on 
another possibility to realize the chiral symmetry breaking in an Abelian gauge theory. 
Suppose that there are monopoles but no electron in an Abelian gauge theory. 
This is equivalent to the QED 
via the dual symmetry, 
so that 
the monopole charge 
is asymptotic non-free. 
Then, let us add electrons 
with an electric charge satisfying the Dirac's quantization condition. 
When the number of electrons 
is much smaller than that of monopoles, 
their effect on the beta function is negligible. 
This implies that 
the monopole charge remains asymptotic non-free 
at least until the monopole charge becomes much smaller than the electron charge. 
Then the Dirac's quantization condition implies that 
the electric charge increases as the energy scale decreases, 
i.e., it is asymptotic free. 
As the electric charge increases, 
its effect on the bata function becomes efficient 
and the beta function may be cancelled and suppressed. 
Then there may be a certain value of electric charge at which the beta function is absent. 
In Ref.~\cite{Akiba:1985rr}, 
the self energy of chiral fermion has been calculated 
and it has been shown that the chiral symmetry breaking occurs 
in an Abelian gauge theory 
when the gauge coupling is 
larger than the threshold value of $e_c = 2 \pi /\sqrt{3}$. 
This implies that 
the Abelian gauge theory can realize the chiral condensation 
when the number of electrons is sufficiently smaller than that of monopoles.

\vspace{1cm}

%
\section*{Acknowledgments}
T.T.Y thanks H. Murayama and Y. Tachikawa for grateful discussions and comments. 
M.Y. thanks W. Buchm\"{u}ller for kind hospitality at DESY where this work was finished. 
This work is supported by Grant-in-Aid for Scientific Research 
from the Ministry of Education, Science, Sports, and Culture
(MEXT), Japan, 
No. 26104009 and No. 26287039 (T.T.Y), 
World Premier International Research Center Initiative
(WPI Initiative), MEXT, Japan,
and the Program for the Leading Graduate Schools, MEXT, Japan (M.Y.).
M.Y. acknowledges the support by the JSPS Research Fellowships for Young Scientists, No. 25.8715.
%

\vspace{1cm}



\end{document}